\documentclass{moriond}

\def\mco{\multicolumn}

\def\be{\begin{equation}}
\def\ee{\end{equation}}
\def\bea{\begin{eqnarray}}
\def\eea{\end{eqnarray}}

\usepackage[percent]{overpic}
\usepackage{tikz}
\usepackage{xcolor}
\usepackage{amsfonts,amsmath}
\usepackage{float,multirow}
\usepackage{xspace}
\def\elivn{\ensuremath{E_{\text{LIV},n}}\xspace}
\def\elivun{\ensuremath{E_{\text{LIV},n=1}}\xspace}
\def\elivde{\ensuremath{E_{\text{LIV},n=2}}\xspace}
\def\eliv{\ensuremath{E_{\text{LIV}}}\xspace}
\def\kjp{\ensuremath{\kappa^\text{JP}(z)}\xspace}%
\newcommand{\Photo}{}

\begin{document}
\vspace*{4cm}
\title{LORENTZ INVARIANCE VIOLATION SEARCH WITH FLARING ACTIVE GALACTIC NUCLEI OBSERVATIONS OF THE FIRST LARGE-SIZED TELESCOPE OF CTAO}

\author{C. PLARD$^1$ AND S. CAROFF$^2$ ON BEHALF OF THE CTAO-LST PROJECT}

\address{
    $^1$LPNHE, 4 place Jussieu, Paris 75252, France\\
    $^2$LAPP, 9 Chemin de Bellevue BP110, Annecy F-74941, France
}

\maketitle\abstracts{
The rapid variability observed in very-high-energy (VHE) sources—such as pulsars, gamma-ray bursts (GRBs), and flares from active galactic nuclei (AGN)—can be used to detect or constrain a potential violation of Lorentz invariance (LIV). These effects can be investigated by measuring time lags in the arrival of VHE photons. However, an important source of uncertainty arises from intrinsic processes within the sources themselves that may induce photon delays unrelated to LIV. To address this challenge, we aim to combine observations of different sources, located at different redshifts. In this study, we present the results of a standardized analysis applied to all AGN observations conducted by the first Large-Sized Telescope of the upcoming Cherenkov Telescope Array Observatory. Our analysis includes a systematic search for intra-night variability in archival data from nights with significant excess detections for target candidates. By combining these observations, we derive constraints on the characteristic energy scales at which LIV deterministic or stochastic effects are expected to manifest.}

\section{Lorentz invariance violation search using delays in gamma-rays arrival times}

The quantized structure of spacetime hypothesized by some quantum gravity models would result in an energy dependence of the velocity of massless particles, including photons, leading to a Lorentz invariance violation (LIV). This is modeled by a modification of the dispersion relation of massless particles in~\cite{amelino_deter}: 
\begin{equation}\label{eq:disp}
E^2 = p^2 c^2 \left[1 \pm \sum_{n=1}^\infty \bigg (\frac{E}{\elivn}\bigg )^n \right],
\end{equation}
where $p$ denotes the photon momentum, $c$ the speed of light in vacuum, the $\pm$ sign corresponds to superluminal or subluminal motion, $E$ is the photon energy, and \elivn is the characteristic energy scale associated with such LIV effects. These effects would manifest as delays in the arrival times at Earth of photons emitted by an astrophysical source at different energies. They may depend universally on the photon energy and are referred to as \emph{deterministic}~\cite{amelino_deter} LIV effects, which are the most studied. The expected delay, in seconds, for two photons $i$ and $j$ of energies $E_i$ and $E_j$ is~\cite{j_p}:
\begin{equation}\label{eq:delay}
\Delta t_n \simeq \pm \frac{n+1}{2}\frac{\Delta E^n}{H_0 \elivn^n} \kappa(z),
\end{equation}
where $\Delta E^n=E_i^n - E_j^n$, $H_0$ is the Hubble constant, and $\kappa(z)$ is a propagation factor that accounts for cosmological expansion and depends on the source redshift $z$. 
Another LIV effect, referred to as \emph{stochastic}~\cite{amelino_stoch}, could also arise. In this case, the photon arrival times at Earth would be normally distributed around the arrival time $T_c$ of low-energy photons (for which quantum gravity effects are negligible, i.e. $v(E) = c$), with a standard deviation $\sigma_t$ given by the same expression as $\Delta t$ (see Equation \ref{eq:delay}). In this case, only the case $\sigma_t >0$ can be probed. 
In the absence of any detected effect, we aim to constrain \elivn, which is expected to approach the Planck scale ($E_P \sim 10^{19}$ GeV), at linear ($n=1$) and quadratic ($n=2$) orders. Suitable sources must emit photons across a broad energy spectrum with sufficient flux and, crucially, exhibit significant fast temporal variability. Pulsars, gamma-ray bursts (GRBs), and flaring active galactic nuclei (AGN) are therefore particularly suitable candidates.

Nevertheless, such studies face two major challenges. The first is that the choice of the $\kappa(z)$ function can significantly impact the LIV sensitivity~\cite{kappa}. In this work, we adopt the most widely used formulation, the "Jacob \& Piran" function~\cite{j_p}, denoted \kjp, which increases with $z$. Using a sample of sources spanning a wide range of redshifts improves the robustness of the analysis with respect to this choice. The second challenge is the possible existence of intrinsic source delays, which may contribute to the observed delay at Earth. These are expected to arise from internal physical processes and may vary between sources or individual AGN flares, while being independent of redshift. In contrast, a delay induced by LIV would appear consistently across all sources and AGN flares, with a redshift dependence governed by the $\kappa(z)$ function. Consequently, combining observations from multiple sources or flares located at different redshifts enables discrimination between these two contributions. This has motivated the creation of a working group~\cite{icrc_glivwg}~\cite{first_combined_study} involving four major Cherenkov telescope experiments: H.E.S.S., MAGIC, VERITAS, and LST-1 of CTAO. In this work, we present the contribution of LST-1 to this effort, based on a consistent analysis of all suitable AGN observations. We developed a pipeline named \texttt{gammaLIV}~\cite{gammaliv}, based on \texttt{gammapy}~\cite{gammapy2023}, the official high-level data analysis framework of CTAO.

\section{Search for time variability within LST-1 observations of flaring AGN}

Our sample consists of all AGN observed by LST-1 from January 2021 to May 2025 with known redshifts. After standard quality selection, the dataset includes 32 sources observed over 507 hours, distributed across 502 observing nights. 

To search for intra-night variability in these data, it is necessary to maximize source detection. To enhance the signal-to-background ratio, event selection criteria, commonly referred to as "cuts", are applied. One such criterion is based on the \emph{gammaness}, a score ranging from 0 to 1 assigned to each reconstructed event, indicating the likelihood that the primary particle is a gamma ray rather than a background event (mainly protons from cosmic rays). The final dataset includes events with a gammaness value above a chosen threshold, which we aim to optimize. For this purpose, we developed a method~\cite{icrc_lst} based on adapting the Crab Nebula signal to mimic a given source, enabling a source-specific optimization of the gammaness cut. We constructed a catalog of optimized cuts for all AGN observed by LST-1, which is available to the LST community for other studies.

In AGN data, the fast temporal variability required for LIV studies is expected on the timescale of a single observing night. The first step is therefore to select nights during which a given source is significantly detected, defined as a total detection significance greater than $5\sigma$. For each of the resulting 161 \emph{significant} nights, corresponding to 9 detected sources, we derived the gamma-ray integrated flux, referred to as the \emph{light curve} (LC). 
A constant function was fitted to each LC, and a night was considered to exhibit intra-night variability if the constant-flux hypothesized was rejected at the $5\sigma$ level (corresponding p-value $<5\times10^{-7}$). Five observing nights of BL Lacertae ($z=0.069$) passed this criterion: August 2nd, 3rd, 8th, and 9th, 2021, and October 20th, 2022. 
As a sanity check, the same method was applied to Crab Nebula data, which is not expected to show intra-night variability, as it is a stable source at TeV energies.

The LIV analysis applied to these five \emph{variable} nights requires a parametrization of both their spectral and temporal distributions, the latter at low energies---that is selecting events with energy below the median of the sample. We evaluated whether a log-parabola model provides a statistically significant improvement, at the $5\sigma$ level, over a simple power-law model for describing the spectrum of each night. All spectra were found to be best described by a power-law model. The LC of each variable night was then fitted with the simplest model—ranging from an affine function to a sum of $n$ Gaussian functions plus a constant—for which the associated p-value exceeded 0.05 (i.e., not excluded at $2\sigma$). The temporal distributions of all the variable nights were described with $n=1$ Gaussian, except for the night of August 8th, 2021, which required $n=3$ Gaussians.

\section{Constraint on LIV deterministic and stochastic effects and discussion}

The search for time delays in these data is performed using a maximum likelihood (ML) method~\cite{ml_method}, implemented in the \texttt{LIVelihood} package~\cite{first_combined_study}, developed for LIV studies with Cherenkov telescopes. The likelihood function $\mathcal{L}(\lambda)$ is maximized with respect to the \emph{lateshift} parameter $\lambda_n$, expressed in s TeV$^{-n}$ and defined as:
\begin{equation}
    \lambda_n = \frac{\Delta t_n\text{ or }\sigma_t}{\Delta E^n\kappa_n(z)}= \pm \frac{n+1}{2H_0}\frac{1}{\elivn^n}.
\end{equation}
For each observing night (flare) or source $s$, the corresponding likelihood $\mathcal{L}_s(\lambda)$ is constructed as the product of the probability density functions describing the detection of each event $i$ with energy $E_i$ and arrival time $t_i$. The combination of multiple sources or nights is obtained by summing the corresponding log-likelihoods, $\sum_s\log\mathcal{L}_s(\lambda)$.
The extraction of the results follows a procedure~\cite{icrc_glivwg}~\cite{icrc_lst} involving simulations, which allow for calibration of the method as well as estimation of biases and systematic uncertainties. 
The total likelihood is fitted to each of one thousand simulated datasets, yielding distributions of reconstructed lateshift values and associated errors at a 95\% confidence level. The final lateshift value and its uncertainties are taken as the means of these distributions.

\begin{figure}
\begin{center}
\begin{overpic}[scale=0.6]{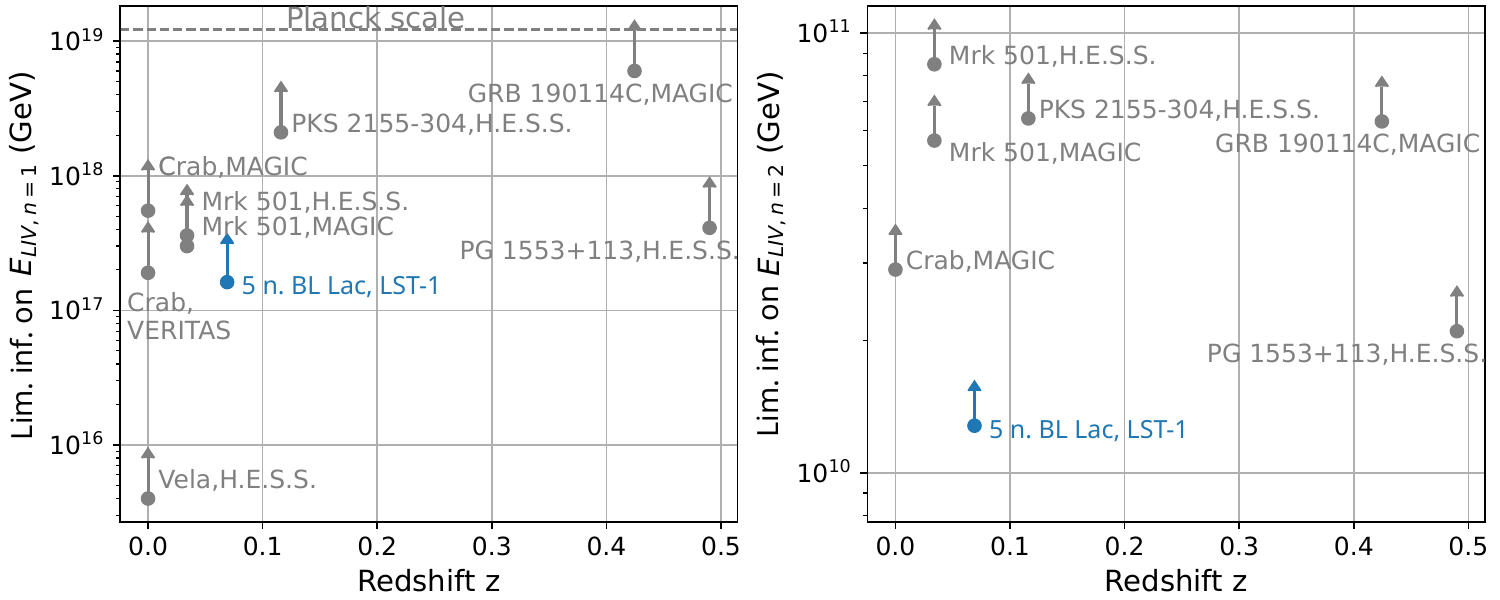}
            \put(35,8){\makebox(0,0){\textcolor{red!60}{\LARGE PRELIMINARY}}}
            \put(85,8){\makebox(0,0){\textcolor{red!60}{\LARGE PRELIMINARY}}}
\end{overpic}
\end{center}
\caption[]{Comparison between limits (subluminal case) obtained from various Cherenkov telescopes~\cite{first_combined_study}  at the linear $n=1$ (left) and quadratic $n=2$ (right) orders.}
\label{fig:lims}
\end{figure}

The results are summarized in Table \ref{tab:lims}: no LIV effect has been observed at either linear or quadratic order, allowing us to constrain \elivn. Figure \ref{fig:lims} presents a comparison of the limits on subluminal deterministic LIV effects with those obtained from other Cherenkov telescope data. Aside from these values, the most stringent constraint to date is~\cite{grb090510_fermi_liv_2013} \eliv $> 9.3 \times 10^{19}$ GeV, derived from Fermi observations of GRB 090510. Although our constraint is one to two orders of magnitude lower, it has been shown~\cite{kappa}~\cite{icrc_glivwg} that it strongly depends on the choice of the $\kappa(z)$ model. Regarding stochastic effects, the only study available in the literature reports~\cite{amelino_stoch} \eliv $> 3.2 \times 10^{19}$ GeV, also based on Fermi data of GRB 090510. Following the suggestion of a previous work~\cite{kappa}, we plan to apply our analysis pipeline to these data, both as a cross-check and to include this source in the combined analysis. Finally, this work represents the first combination of multiple AGN flares and is therefore the most robust with respect to potential intrinsic source delays. This dataset is intended to be combined with observations from other facilities, further improving robustness against both astrophysical effects and propagation modeling through $\kappa(z)$.
\begin{table}[H]
\caption[]{Lateshift observed in data and the corresponding constraint on the characteristic energy of LIV effects.}
\vspace{0.4cm}
\begin{center}
\begin{tabular}{|l|l|l|l|}
\hline
\mco{2}{|c|}{\textcolor{red!60}{\Large PRELIMINARY}} & Deterministic LIV & Stochastic LIV \\\hline
\rule[-1.5ex]{0pt}{4ex}\multirow{2}*{$n=1$} & $\lambda_{n=1}$ (s TeV$^{-1}$) & $-58.54^{+ 2756.74}_{- 2951.64}$ & $2952^{+ 7799}_{-N/A}$ \\
& sublum. \elivun $>$ & 1.62 $\times 10^{17}$ GeV & 5.30 $\times 10^{16}$ GeV\\
& superlum. \elivun $>$ & 1.45 $\times 10^{17}$ GeV& \\\hline
\rule[-1.5ex]{0pt}{4ex}\multirow{2}*{$n=2$} & $\lambda_{n=2}$ (s TeV$^{-2}$) & $-669.27^{+4644.88}_{-5494.87}$ & $1517^{+13558}_{- N/A}$ \\
& sublum. \elivde $>$ & 1.28 $\times 10^{10}$ GeV& 6.59 $\times 10^{9}$ GeV\\
& superlum. \elivde $>$ & 1.03 $\times 10^{10}$ GeV&  \\
\hline
\end{tabular}
\end{center}\label{tab:lims}
\end{table}

\section{Conclusion}

We performed a systematic and consistent analysis of all AGN with known redshift observed by LST-1 up to May 2025, searching for rapid variability in the LCs of individual observing nights. Five nights of BL Lac exhibited significant intra-night variability and were combined to search for potential deterministic or stochastic LIV effects. As no significant signal was detected, lower limits on $E_{LIV,n}$ were derived at linear ($n=1$) and quadratic ($n=2$) orders. This constitutes the first constraint obtained from a combination of AGN flares observed by a Cherenkov telescope. Future work includes incorporating additional sources into the combined analysis, starting with variable nights detected at a lower variability threshold from LST-1, and extending this systematic approach to AGN observations from other telescopes in the working group, namely H.E.S.S., MAGIC, and VERITAS.

\section*{Acknowledgments}

This work was conducted in the context of the CTAO-LST project. The corresponding authors gratefully acknowledge financial support from the following agencies and organisations: \url{https://www.ctao.org/for-scientists/library/acknowledgments/} and the funding from the French Programme d'investissements d'avenir through the Enigmass Labex.

\section*{References}
\bibliography{moriond}


\end{document}